\documentclass[11pt]{article}
\usepackage{amssymb,amsmath,graphicx,epsfig} 
\def\ltap{\raisebox{-.4ex}{\rlap{$\sim$}} \raisebox{.4ex}{$<$}} 
\def\gtap{\raisebox{-.4ex}{\rlap{$\sim$}} \raisebox{.4ex}{$>$}} 

\def\issue(#1,#2,#3){{\bf #1}, #2 (#3)} 

\def\APP(#1,#2,#3){{\rm Acta Phys.\ Polon.} \ \issue({\bf #1},#2,#3)}
\def\ANP(#1,#2,#3){{\rm Annals of Physics} \ \issue({\bf #1},#2,#3)}
\def\ARNPS(#1,#2,#3){{\rm Ann.\ Rev.\ Nucl.\ Part.\ Sci.} \ \issue({\bf #1},#2,#3)}
\def\CPC(#1,#2,#3){{\rm Comp.\ Phys.\ Comm.} \ \issue({\bf #1},#2,#3)}
\def\CIP(#1,#2,#3){{\rm Comput.\ Phys.} \ \issue({\bf #1},#2,#3)}
\def\EPJ(#1,#2,#3){{\rm Eur.\ Phys.\ J.} \ \issue({\bf #1},#2,#3)}
\def\EPJD(#1,#2,#3){Eur.\ Phys.\ J. Direct\ C \ \issue({\bf #1},#2,#3)}
\def\IJMP(#1,#2,#3){{\rm Int.\ J.\ Mod.\ Phys.} \ \issue({\bf #1},#2,#3)}
\def\JHEP(#1,#2,#3){{\rm J.\ High Energy Physics} \ \issue({\bf #1},#2,#3)}
\def\JP(#1,#2,#3){{ J.\ Phys.} \ \issue({\bf #1},#2,#3)}
\def\MPL(#1,#2,#3){{Mod.\ Phys.\ Lett.} \ \issue({\bf #1},#2,#3)}
\def\NP(#1,#2,#3){{Nucl.\ Phys.} \ \issue({\bf #1},#2,#3)}
\def\NIM(#1,#2,#3){{ Nucl.\ Instrum.\ Meth.} \ \issue({\bf #1},#2,#3)}
\def\PL(#1,#2,#3){{ Phys.\ Lett.} \ \issue({\bf #1},#2,#3)}
\def\PR(#1,#2,#3){{ Phys.\ Rev.} \ \issue({\bf #1},#2,#3)}
\def\PRL(#1,#2,#3){{ Phys.\ Rev.\ Lett.} \ \issue({\bf #1},#2,#3)}
\def\SJNP(#1,#2,#3){{ Sov.\ J. Nucl.\ Phys.} \ \issue({\bf #1},#2,#3)}
\def\ZP(#1,#2,#3){{Zeit.\ Phys.} \ \issue({\bf #1},#2,#3)}
\def\beq{\begin{equation}}
\def\eeq{\end{equation}}
\def\barr{\begin{array}}
\def\earr{\end{array}}

\def\gev{\, {\rm GeV}}
\def\mev{\, {\rm MeV}}
\def\ev{\, {\rm eV}}

\def\bea {\begin{eqnarray}}
\def\eea {\end{eqnarray}}

\def\ket {\rangle}

\def\minos{{\sc minos}\ }

\begin{document}
\begin{center}
{\Large {\bf 
{Mutual consistency of the MINOS and MiniBooNE Antineutrino Results
and Possible CPT Violation}}}
\\[5mm]
\bigskip
{\sf Debajyoti Choudhury} $^a$,
{\sf Anindya Datta} $^b$, and
{\sf Anirban Kundu} $^b$
\bigskip

$^a$ {\footnotesize\rm
Department of Physics and Astrophysics,
University of Delhi, Delhi 110 007, India}

\bigskip

$^b$ {\footnotesize\rm 
Department of Physics, University of Calcutta, \\
92, Acharya Prafulla Chandra Road, Kolkata 700 009, India}

\normalsize
\vskip 10pt

{\large\bf Abstract}
\end{center}

\begin{quotation} \noindent 

Discussing the recent MINOS data on $\bar\nu_\mu$ disappearance and
the MiniBooNE data on $\bar\nu_\mu \to \bar\nu_e$ oscillation, we show
that the while the respective best fits are inconsistent with each
other, significant overlap of allowed regions does exist. Assuming
only three neutrino species, the data indicates a discrepancy of mass
levels and mixing angles between the neutrino and the antineutrino
sectors.  We show that the existing data can be reconciled with a
model of explicit CPT violation in the neutrino sector and estimate
the magnitude of the required violation.

\vskip 10pt
PACS numbers: {\tt 11.30.Er, 14.60.Pq}\\
\end{quotation}
\begin{flushleft}\today\end{flushleft}

\noindent
Experimental evidence in support of tiny but finite neutrino masses
demands an extension of the Standard Model (SM) \cite{parke}. 
Recent results from
the \minos \cite{MINOS} and MiniBOONE \cite{MiniBOONE}
collaborations not only re-assert this, but also force us to think
about violation of one of the cornerstones of theoretical physics,
namely the CPT theorem. The \minos collaboration has looked for, and
found, signals of oscillations in both $\nu_\mu$ and $\bar \nu_\mu$
beams. Perhaps their most surprising finding is that parameters
governing $\bar\nu_\mu$ disappearance are different from those
governing $\nu_\mu$.  The most straightforward interpretation, namely
that the neutrinos have masses different from those of anti-neutrinos,
runs counter to the CPT theorem, which, in turn, is a good symmetry
for any local field theory defined in a Minkowski space-time. This is
probabaly the first time in the history of physics that we are faced
with a situation where to explain a certain experimental result,
violation of CPT is asked for.

Yet another longstanding anomalous result is that from the LSND
experiment~\cite{lsnd_ref}
which claimed an evidence for $\bar \nu_e$ oscillating into
something. The key issue was that the mass scale was quite distinct
from either of those deemed to be responsible for the solar neutrinos
(ostensibly $\nu_e \leftrightarrow \nu_\mu$) or the atmospheric ones
($\nu_\mu \leftrightarrow \nu_\tau$). While this could be explained by
postulating a sterile neutrino, this explanation is highly disfavoured
by a global fit of all neutrino data \cite{neu_global}, in particular
the SNO neutral current measurements \cite{SNO}.  To resolve this anomaly,
various experiments have been performed. The latest in this 
line is MiniBooNE, 
which has claimed consistency with the LSND results.  It should be
noted at this point that the LSND result could, in principle, be
explained in the framework of CPT violation~\cite{lsnd_cpt}.

In view of these two rather startling results, each demanding a
drastic step away from the SM, it is worthwhile to consider the
possibility of explaining both within a single framework (whether CPT
violation or otherwise). However, before delving into this, it would
be prudent to examine the (degree of) consistency of the two
experiments with each other, and this is what we begin with.

At this point, it would be wise to remember that, matter being CP
asymmetric, $\nu$'s and $\bar\nu$'s, while propagating through it,
have differing effective masses. Assuming that they can have only SM
interactions (and keeping in mind that the $1 \leftrightarrow 3$ mixing
 is small), the
matter effect is negligible in $\nu_\mu - \nu_\tau$ sector. However,
the inclusion of a non-standard interaction (NSI) involving $\nu_\mu$
and $\nu_\tau$ can offset this result, and the consequent matter
effect can be substantial even for the relatively small \minos
baseline. Ref\cite{NSI} seeks to exploit this to explain the \minos
result. However, two subtle issues need to be considered. In any
realistic NSI framework, allowing for a flavour violating coupling
involving $\nu_\mu$ and $\nu_\tau$ necessarily implies a similar
interaction involving $\mu$ and $\tau$, thereby, potentially
triggering the so far unobserved decay $\tau \rightarrow \mu \gamma$.
Furthermore, any attempt to explain the MiniBOONE results within the
same NSI framework would, generically, lead to a larger flavour
violation involving $e$ and $\mu$ (on account of the larger mass
difference observed), resulting in very large $\mu \to e
\gamma$. Note, however, that the MiniBooNE $\bar\nu_e$ excess (albeit
in a different energy range) has also been ascribed to an
underestimation of SM-induced single-photon background
events \cite{Hill:2010zy}.
 
The Pontecorvo-Maki-Nakagawa-Sakata (PMNS) matrix\cite{PMNS} that relates 
the neutrino mass $\nu_i \, (i = 1 \cdots 3)$ 
and  flavour $\nu_\alpha \, (\alpha = e, \mu,\tau)$ eigenstates
through 
\(
|\nu_\alpha\ket = \sum_{i=1}^3 U_{\alpha i}^\ast |\nu_i\ket\, ,
\)
is usually parametrized as 
\beq
U = U_{23}(\theta_{23}) \, U_{13}(\theta_{13}) \, U_{12}(\theta_{12})
    \label{eq:PMNS}
\eeq
where $U_{ij}$ represents an orthogonal rotation in the $ij$ plane
through the angle $\theta_{ij}$. 
In the above, (1,2) is the solar pair, with $m_2$ being 
marginally larger than $m_1$. As for the atmospheric pair, $m_3 > m_2$ denotes
the normal hierarchy (NH), while $m_3 < m_2$ implies the inverted one (IH). 
For our numerical analysis, we use \cite{pdg}
\beq
\barr{rclcrcl}
\Delta m_{21}^2 & = & (7.59\pm 0.20)\times 10^{-5} \ev^2
& \quad & \sin^2(2\theta_{12}) & = & 0.87 \pm 0.03\,
\\
|\Delta m_{23}^2| & = & (2.43\pm 0.13)\times 10^{-3} \ev^2\
& & 
\sin^2(2\theta_{23}) & = & 1.0
\earr
   \label{nu-param}
\eeq
where $\Delta m_{ij}^2 \equiv m_i^2 - m_j^2$. 
In eq.\ (\ref{eq:PMNS}), we have neglected 
the Majorana phases (which are irrelevant for the discussion at hand) 
and have also assumed that there is no CP-violating phase, an assumption 
that is not a drastic one in view of the fact that $\theta_{13}$ is 
constrained to be very small ($\theta_{13} \ltap 7^\circ$).

Henceforth, we shall adopt the convention whereby
lowercase letters ($m$ and $\theta$) are used for the $\nu$'s, and
uppercase letters ($M$ and $\Theta$) for $\bar\nu$'s.  Note that
eq.\ (\ref{nu-param}) applies only to the $\nu$'s. Furthermore, if CPT is
violated, there is no way to relate the $nu$ hierarchy with the
$\bar\nu$ one. A constraint that binds the two sector 
is the cosmological one on
$m^{\rm tot} \equiv \sum m_\nu$, with the sum ranging  over 
SM-like $\nu$'s and $\bar\nu$'s. Current data restricts 
$m^{\rm tot} \le 0.56 \ev $ if the flat $\Lambda$CDM model is 
assumed and $m^{\rm tot} \le 0.94 \ev $ if a generic 
dark energy source is considered~\cite{thomas_abdalla_lahav}.

\minos looks at $\bar{\nu}_\mu$ disappearance, so technically it can go
to either $\bar{\nu}_e$ or $\bar{\nu}_\tau$.
Assuming $\bar\nu_\mu\to \bar\nu_\tau$, \minos obtained results
in the antineutrino 2-3 sector which are slightly different from the 
corresponding data on neutrinos:
\beq
\Delta M^2_{23} = (3.36^{+0.45}_{-0.40} \pm 0.06)\times 10^{-3} \ev^2 \, ,\ \ 
\sin^2(2\Theta_{23}) = 0.86\pm 0.11 \pm 0.01\,,
\eeq
where the first error is statistical and the second is systematic. 
Thus, there is more than $2\sigma$ discrepancy between neutrino and 
antineutrino data.

\begin{figure}[htbp]
\vspace{-10pt}
\centerline{\hspace{-3.3mm}
\rotatebox{0}{\epsfxsize=7cm\epsfbox{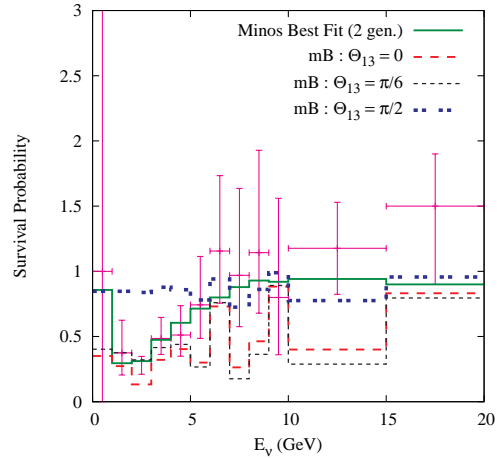}}}
\vspace{-15ex}
\caption{Expected survival probability at \minos for different $\Theta_{13}$
if the MiniBooNE best fit is to be incorporated in a 3-generation scheme.}
\label{fig:mBosc}
\end{figure}
Considering the appearance of $\bar \nu_e$ in a $\bar\nu_\mu$ 
beam and analysing the counts for quasi-elastic
scattering for antineutrino energies on the range $475 \mev \leq E_\nu^{QE}
\leq 3 \gev$, MiniBooNE claims evidence for oscillations with parameters 
distinctly different from the solar ones. Admittedly, the error bars 
are yet very large, but the best fit, viz.
\beq
\Delta M^2_{12} = 0.064~{\rm eV}^2\,,\ \ 
\sin^2(2\Theta_{12}) = 0.96\,,
\eeq
also represents virtually the smallest value for $\Delta M^2_{12}$.
Together, the two experiments suggest the need for at least 5
neutrino/antineutrino levels.  Moreover, the neutrinos should be Dirac
particles\footnote{Majorana fermions can be CPT violating, but here we
need different mass values.}. However, it is imperative to check
whether \minos and MiniBooNE findings are mutually consistent. Note
that either analysis is contingent on an effective two-generation
hypothesis. Whereas this is a fairly good assumption in the neutrino
sector (the solar splitting does not materially affect the atmospheric
oscillation), for the antineutrinos it might not be so, as the lowest
possible value (at 99\% CL) of $\Delta M^2_{12}$ claimed by
MiniBooNE is about one order of magnitude larger than the \minos value
of $\Delta M^2_{23}$.  The \minos survival probability $P$ 
reaches a plateau ($P \sim 1$) for $E_\nu \gtap 8$ GeV. If we
incorporate MiniBooNE as well, a three-flavour analysis needs to be
performed, and consequently, large oscillations in $P$ survive
into the high-$E_\nu$ region, and even after bin-averaging, the
resultant $P$ is inconsistent with \minos (see Fig.\ref{fig:mBosc}).
Evidently, non-zero values of $\Theta_{13}$ do not help reconcile the
two, and, in addition, the agreement worsens in the small $E_\nu$
region as well.

\begin{figure}[htbp]
\vspace{-24pt}
\centerline{\hspace{-3.3mm}
\rotatebox{0}{\epsfxsize=7cm\epsfbox{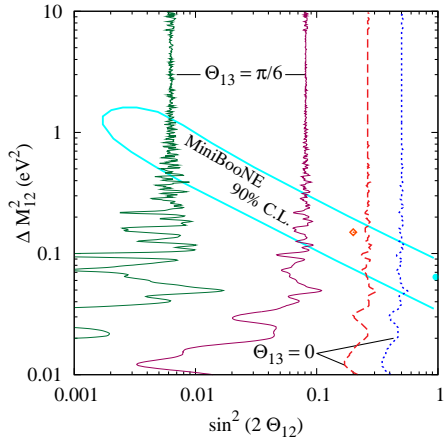}}}
\vspace{-19ex}
\caption{Parameter space consistent with \minos at 
90\% (left) and 95\% C.L. (right) for two values of $\Theta_{13}$. 
Also shown is the 90\% CL contour of MiniBooNE. The point on the right 
edge is the MiniBooNE best fit and the diamond is our benchmark point BP2.}
\label{fig:chisq}
\end{figure}

\begin{figure}[htbp]
\vspace{-24pt}
\centerline{\hspace{-3.3mm}
\rotatebox{0}{\epsfxsize=7cm\epsfbox{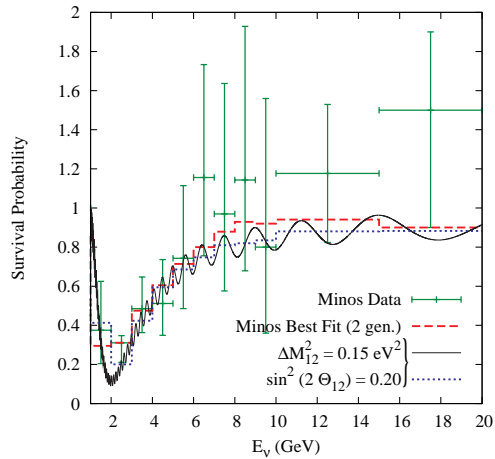}}}
\vspace{-18ex}
\caption{$\bar\nu_\mu$ survival probability at \minos for BP2.
}
\label{fig:benchmark}
\end{figure}

This disagreement is omnipresent over the entire $1\sigma$
allowed parameter space for MiniBooNE and, at this level, the two
experiments are clearly inconsistent with each other.  The 90\% CL
allowed regions do overlap though (Fig.\ref{fig:chisq}). That the 
overlap decreases with 
increasing $\Theta_{13}$ can be understood by realising that 
a large $\Delta M^2_{12}$ implies a large  $\Delta M^2_{13}$, and 
coupled with a non-zero  $\Theta_{13}$, it sets off additional 
oscillations in $P$. The
diamond in Fig.\ref{fig:chisq} indicates one of our benchmark
points; we show, in Fig.\ref{fig:benchmark}, the corresponding 
survival probability as well as its bin-averaged value. 
It is evident that while this (and similar others) choice of parameters
is consistent with \minos as of now, 
an improvement in the \minos energy resolution would go a long way 
in confirming (or ruling out) this solution. 
Also note that the benchmark point is 
allowed by the Bugey and Karmen experiments too.

CPT violation in the neutrino sector has been considered earlier
\cite{glashow} and can be classified mainly into two categories.  One
class of such field theories are non-local. The others are local, and,
generically can be represented as an interaction term often
expressible in terms of a preferred direction.
However, it has been shown \cite{green} that CPT
violating theories necessarily do not respect Lorentz
symmetry. Possible mechanisms to violate CPT have been discussed by
several authors \cite{kostelecky}. Consequences of such CPTV
interactions on neutrino oscillation have been extensively studied in
the literature \cite{CPT_more}.

Rather than consider a particular model, we adopt
a purely phenomenological approach. 
Given that the neutrinos and
their CPT partners seem to have differing masses, 
we assume that the only relevant CPT violation 
appears in the effective (anti-)neutrino mass matrices. 
Apart from restricting us to
Dirac neutrinos, the rest of the analysis is independent of
any particular framework proposed in the literature. 
The mass matrices, in the flavour basis, can then be parametrized as 
\beq
m^{\rm flav}_{ij} = \mu_{ij} + \epsilon_{ij}\, , \quad
M^{\rm flav}_{ij}  = \mu_{ij} - \epsilon_{ij}
\eeq
where $\mu$ ($\epsilon$) are the CPT conserving (violating) parts 
respectively. These would then be diagonalized 
by the respective PMNS matrices $U(\theta_{ij})$ and $\bar{U}(\Theta_{ij})$,
assuming the same diagonalizing matrices for left- and right-chiral
$\nu$ and $\bar\nu$ fields.

With too many parameters and too little data, we adopt the constraint
that the lightest of neutrino and antineutrino are exactly degenerate.
Still, given that the signs of $\Delta M^2_{ij}$ are unknown, as many
as twelve hierarchy combinations are possible: $N_{ijk}$ and $I_{ijk}$
where $N$ ($I$) refer  to the normal (inverted) hierarchy in the
$\nu$ sector while $ijk$ indicate the hierarchy $M_i < M_j < M_k$.
Rather than scan over the parameter space, we choose to discuss the
numerical results for two specific benchmark points.  For both,
$\Delta M_{23}^2$ and $\sin^2 2\Theta_{23}$ are held to the \minos
central value, while $\theta_{13}=\Theta_{13}=0$ (as Fig
\ref{fig:chisq} demonstrates, a non-zero $\Theta_{13}$ tends to worsen
the fit).  BP1 ignores the MiniBooNE result altogether, and holds
$\Delta M_{12}^2 = \Delta m_{12}^2$ and $\Theta_{12}= \theta_{12}$.
BP2, on the other hand, refers to the point illustrated in
Figs.\ref{fig:chisq}\&\ref{fig:benchmark}.  As can be expected, the
results drastically depend on this choice. Of particular importance is
the cosmological constraint on $m^{\rm tot}$, which is 
rather severe on BP2, and would effectively rule out
the allowed parameter points further away from the MiniBooNE best fit.

\begin{table}[h]
\begin{center}
\begin{tabular}{|c|c|c|c|c|c|}
\hline
& Scheme & $R_1 < 1$ & $R_2 < 1$ 
        & \multicolumn{2}{|c|}{${\rm Max}(m_{\rm lowest})$ (meV)} \\
\cline{5-6}
& &   &  & $0.56 \ev$ & $0.94 \ev$ \\
\hline
BP1   &  $N_{123}$, $N_{213}$ &   $\surd$  & $\surd$ &  32  & 76\\ 
      & $N_{312}$, $N_{321}$ &   $\surd$  &         &  28  & 73 \\
      &  $I_{123}$, $I_{213}$ &   $\surd$  &         &  86  & 153\\
      &  $I_{312}$, $I_{321}$ &   $\surd$  & $\surd$ &  83  & 151\\
\hline
BP2   & $N_{123}$, $N_{132}$ & $\surd$ &    & --- & 4.8   \\ 
      & $N_{231}$, $N_{321}$ &   &     & 1.5 & 34   \\ 
      & $I_{123}$, $I_{132}$ &$\surd$&   & --- & 2.8     \\ 
      & $I_{231}$, $I_{321}$ & $\surd$& $\surd$ ($I_{321}$)  & 7.2 &  101    \\ 
\hline
\end{tabular}
\caption{Allowed neutrino and antineutrino hierarchies, 
and the associated measures of CPTV (see text).}
\label{tab:hier}
\end{center}
\end{table}

Not all hierarchies are allowed for a given benchmark point.  While
some are ruled out by dint of the values of $\Delta M^2_{ij}$, others
are strongly disfavoured by cosmological
constraints. We show the maximum possible mass of the lowest $\nu$ (and
hence $\bar\nu$) level consistent with the cosmological bound in the 
last two columns of Table \ref{tab:hier}.
Note that the allowed schemes 
essentially come in pairs.  Within a pair, the quantitative
features are almost identical, a consequence of the fact that they are
related by the flip of the pair of $\bar \nu$'s with the smaller mass
splitting.  (It should be realized here that while more leeway 
may be bought by relaxing the assumption
of lowest-level degeneracy, Table \ref{tab:hier} encapsulates 
the main physical features.)

Clearly, to achieve some of these hierarchies, 
the size of the CPTV parameters $\epsilon_{ij}$ need to be 
relatively large. In the absence of any theory of the same, 
one cannot formulate a precise definition of 
this largeness or the naturalness thereof. 
To this end, we propose two measures of CPT violation, namely,
\beq
R_1 = \frac{{\rm max} (|\epsilon_{ij}|) } {{\rm max}(|\mu_{ij}|)}\,,\ \ 
R_2 = {\rm max} \left( \left| \frac{\epsilon_{ij}} {\mu_{ij}}\right| \right)\,, 
\eeq
If it is some
underlying symmetry that keeps the $\mu_{ij}$ small, we would, naively,
expect it to suppress the corresponding $\epsilon_{ij}$ too and, thus, 
from a model-builder's standpoint, those models where both $R_1$ and $R_2$
are less than unity are a bit more favoured. In Table.\ref{tab:hier}, we 
also display the $R_{1,2}$ properties of the various hiearchies. As 
expected, rather than the normal-inverted ($\nu, \bar\nu)$ and
the inverted-normal cases, it is the normal-normal and inverted-inverted 
hierarchies that are associated with relatively smaller $\epsilon_{ij}$,
with the $I_{321}$ combination being the best from this standpoint.

Several other features are worth noting. 
For BP1, the splitting between $\nu_1(\bar\nu_1)$ and $\nu_2(\bar\nu_2)$
is small compared to $\Delta m^2_{23} (\Delta M^2_{23})$. Consequently, 
there is an approximate $1 \leftrightarrow 2$ symmetry (compared to level 3) 
and we expect 
\(
\epsilon_{13}/\mu_{13} \approx \epsilon_{23}/\mu_{23}
\) 
which is indeed satisfied to a great accuracy. 
For BP2 though, the situation is more complicated, and it is only for the 
$I_{123}$ and $I_{132}$ hierarchies that a similar relation, viz. 
\( \epsilon_{22}/\mu_{22} \approx \epsilon_{33}/\mu_{33} \)
can be found. The large extent of the MiniBooNE-allowed parameter 
space thus points to the difficulty in identifying 
underlying textures for $\epsilon_{ij}$ 
parameters. For example, even for $I_{321}$ alone, the point BP1 is 
consistent with ($\epsilon_{12}, \epsilon_{13} = 0, \, \epsilon_{22} \approx
\epsilon_{11}$), whereas 
BP2 prefers ($\epsilon_{i3} = 0, \, \epsilon_{11} \gg \epsilon_{12}, \epsilon_{22}$). 
To summarise, each of the $\bar\nu$ disappearance results of the \minos
far detector and the recent anomalous MiniBooNE results on $\bar
\nu_\mu \to \bar \nu_e$ individually argues strongly for the
oscillation parameters in the $\bar\nu$ sector to be significantly
different from those in the $\nu$ sector.  Taken together, though, the
two sets of derived parameters are shown to be inconsistent with each
other at the 68\%C.L.; this is because
the large values of $\Delta {\bar M}^2_{12}$
and $\Theta_{12}$, as preferred by MiniBooNE would induce large
oscillations in $P(\bar\nu_\mu\to \bar\nu_\mu)$ for large $E_{\bar
\nu}$ whereas \minos sees a saturating behaviour. However, at 90\%C.L.,
the two experiments turn out to be mutually consistent. It is worth
noting that an improvement of \minos energy resolution at large
$E_{\bar \nu}$ as well the accumulation of more statistics would lead
to a much finer probe of this overlap.

Assuming consistency, the two experiments together present a very
strong argument for properties of $\bar\nu$'s to be radically
different from those of the $\nu$'s, to the extent of violating CPT
invariance.  We explore the possibility that the effective $\nu$-- and
$\bar\nu$--mass matrices differ on account of a CPT violating
interaction.  If CPT is indeed violated, there is no necessity of
having identical hierarchies for $\nu$'s and $\bar\nu$'s. Of the
twelve possible hierarchies consistent with the $\nu$-sector
measurements, only some are found to be consistent with the $\bar\nu$
measurements as well as the cosmological constraints. Furthermore, if 
we demand that the (natural) condition that $\epsilon_{ij}$, 
the CPTV contribution 
to the mass matrix,   be smaller than the the CPT
conserving part, the choices get restricted even further. Finally, 
various textures for $\epsilon_{ij}$ are possible. 

\centerline{\bf{Acknowledgements}}

AD and AK are supported by the 
CSIR, India and the DRS programme of the UGC. 
DC and AK acknowledge the hospitality of the Abdus Salam International 
Centre for Theoretical Physics, Trieste, where a large part of the work was 
completed.


\end{document}